\documentclass[12pt]{article}
\usepackage{graphicx}
\usepackage{amsmath,amssymb,amsfonts}
\graphicspath{ {c:/Users/hacyan/Documents/} }

\begin{document}

\title{Quantum non-linear effects in colliding light beams and interferometers}

\author{S. Hacyan
}

\renewcommand{\theequation}{\arabic{section}.\arabic{equation}}

\maketitle
\begin{center}

{\it  Instituto de F\'{\i}sica,} {\it Universidad Nacional Aut\'onoma de M\'exico,}

{\it Apdo. Postal 20-364, Cd. M\'exico 01000, Mexico.}

\bigskip

e-mail: hacyan@fisica.unam.mx

\end{center}
\vskip0.5cm

\begin{abstract}

\noindent Small corrections to the electromagnetic field in colliding light beams are evaluated taking into account the interaction of light with the quantum vacuum, as predicted by quantum electrodynamics.  Possible implications for very energetic light beams and the radiation pressure in interferometers  (e.g., gravitational waves detector) are considered. It is also shown that a head-on collision produces secondary waves along conical directions at angles $\arccos (1/3)$.

\end{abstract}


\bigskip

Key words: nonlinear electrodynamics; interferometry; quantum vacuum; paraxial approximation

 \maketitle

\newpage
\section{Introduction}

In 1936, Euler and Heisenberg \cite{eh} proved that photons can interact with photons through the intermediate creation of electron-positron pairs, thus introducing nonlinear corrections to the Maxwell equations.
This interaction, being of second order in the fine structure constant $\alpha$, is difficult to observe in practical conditions. Nevertheless, there have been several  works  devoted to the small but not entirely negligible effects predicted by quantum electrodynamics. Some interesting effects are the induction of dichroism and birefringence in the quantum vacuum \cite{klni2, klni,hehe,zava,cade}, the splitting of photons  \cite{ankr,adl,bibi,pimi}, the bending of light by strong magnetic fields  \cite{desv,yole} and other processes related to the propagation of light  \cite{mcpl, gidi,digi,obu,mema,kike}.

The aim of the present paper is to study the possible effects of the quantum vacuum on interacting collimated light beams. In an interferometer, for instance, photons bounce back and forth  (multiple times in a Fabri-P\'erot \cite{ligo2012}) and, if the light beam is sufficiently powerful, there might be small corrections to the radiation pressure and other physical parameters. In particular, having in mind extremely powerful lasers as those used in gravitational waves observatories, it is worth evaluating wether or not the effects of nonlinear electrodynamics should be taken into account among the many possible sources of noise. Fortunately, as shown in the present paper, this is not yet the case, but it may be relatively important in the future if much more powerful light beams were available. In the meantime, some curious properties of light interacting with the quantum vacuum can be deduced theoretically.

The calculations in the present paper are based on the formulation of McKenna and Platzman  \cite{mcpl}, which we review in Section 2 for the sake of completeness.  In Section 3, the nonlinear effects of the interaction between emitted and reflected waves in an interferometer are evaluated within the paraxial approximation. A brief discussion of the results is given in Section 5.

\section{Nonlinear electrodynamics}

As shown in 1964 by  McKenna and Platzman  \cite{mcpl}, the equations of electrodynamics that follow to order $\alpha^2$ from the Euler-Heisenberg Lagrangian are equivalent to the  standard Maxwell equations with ``sources'' arising from the nonlinear terms. Explicitly, the ``density'' and ``current'' are defined as
\begin{equation}
\rho =- \frac{\lambda}{4\pi} \nabla \cdot \boldsymbol{\delta E}
\end{equation}
\begin{equation}
{\bf J} = \frac{\lambda}{4\pi} \Big[ \frac{\partial}{\partial t} \boldsymbol{\delta E} -\nabla \times \boldsymbol{\delta
 B}\Big],
\end{equation}
where the nonlinear terms are
$$
\boldsymbol{\delta E}= 2({\bf E}^2 -{\bf B}^2){\bf E} +7 ({\bf E}\cdot{\bf B}){\bf B}
$$
\begin{equation}
\boldsymbol{\delta B}= 2({\bf E}^2 -{\bf B}^2){\bf B} -7 ({\bf E}\cdot{\bf B}){\bf E}\label{d}.
\end{equation}
Here
$$
\lambda = \frac{\alpha^2}{45\pi m_e^4},
$$
$m_e$ is the electron mass, and naturalized Gaussian cgs units (i.e., $\hbar = c=1$ and $\alpha = e^2$) are used.

The corrections to the electromagnetic field can be expressed in the form
\begin{equation}
{\bf E} ={\bf E_0} + \lambda {\bf E_1} \quad {\bf B} ={\bf B_0} + \lambda {\bf B_1} ,
\end{equation}
where ${\bf E_0}$ and ${\bf B_0}$ are the solutions of the standard linear equations. Thus, the following first order
equations are obtained \cite{mcpl}:
\begin{equation}
\Big(\nabla^2 -\frac{\partial^2}{\partial t^2}\Big){\bf E_1} ({\bf r},t)=4\pi \Big[\frac{\partial}{\partial t}
{\bf J_0}({\bf r},t) + \nabla \rho_0 ({\bf r},t)\Big],
\end{equation}
\begin{equation}
\Big(\nabla^2 -\frac{\partial^2}{\partial t^2}\Big) {\bf B_1} ({\bf r},t)=-4\pi \nabla \times {\bf J_0}({\bf r},t)
,\label{om}
\end{equation}
where it is understood that $\rho_0$ and ${\bf J_0}$ are given in terms of the linear field through Eqs. \eqref{d}.

\section{Linear and nonlinear fields}

In order to fix the ideas, let us consider the incident and reflected waves in an interferometer.
Consider first a standard background field without nonlinear corrections. A stationary polarized EM wave
interfering in vacuum with its reflection from a mirror has the form:
\begin{equation}
{\bf E}({\bf r}, t) = e^{-i\omega t} U({\bf r}) \Big(e^{ikz}+ r e^{-ikz}\Big)~{\bf e_x} \label{E}~,
\end{equation}
 where $r$ is the
(complex) reflection coefficient ($|r| = 1$ for the head-on collision of two identical beams). In the paraxial approximation
$$
\Big| k \frac{\partial U}{\partial z} \Big | \gg \Big|  \frac{\partial^2 U}{\partial z^2} \Big |,
$$
$k=\omega$, and we have
\begin{equation}
{\bf B} = (i\omega)^{-1} \nabla \times {\bf E} = e^{-i\omega t} U({\bf r}) \Big(e^{ikz}- r e^{-ikz}\Big)~{\bf
e_y}.
\end{equation}
Therefore
$$
{\bf E}\cdot{\bf B}=0
$$
\begin{equation}
{\bf E}^2 -{\bf B}^2=4r e^{-2i\omega t} ~U^2({\bf r}).\label{ebeb}
\end{equation}

In the following, we will use the  electromagnetic wave in the simple form (in cylindrical coordinates)
\begin{equation}
U({\bf r})= \mathcal{E}_0 e^{ - \rho^2/w_0^2},
\end{equation}
where $\rho = \sqrt{x^2+y^2}$ and $\mathcal{E}_0$ the electric field amplitude. This  corresponds to the limiting case of a Gaussian beam with width $w_0$ and Raleigh range tending to infinity.

\subsection{Nonlinear corrections}

Let us take \eqref{E} as the electric field inside the interferometer. For the nonlinear terms, we have from \eqref{d} and \eqref{ebeb}
\begin{equation}
\boldsymbol{\delta E} =8r e^{-3i\omega t} U^3 \Big(e^{ikz}+ r e^{-ikz}\Big){\bf e_x}
\end{equation}
\begin{equation}
\boldsymbol{\delta B} =8r e^{-3i\omega t} U^3 \Big(e^{ikz}- r e^{-ikz}\Big){\bf e_y}~,
\end{equation}
and therefore
$$
\nabla \cdot \boldsymbol{\delta E} =0
$$
$$
\frac{\partial}{\partial t} \boldsymbol{\delta E} -\nabla \times \boldsymbol{\delta
 B}=-16 i r \omega e^{-3i\omega t} U^3 \Big(e^{ikz}+ r e^{-ikz}\Big){\bf e_x},
$$
keeping only terms of order $\omega$, still in the paraxial approximation. Thus, Eqs. \eqref{om} are
\begin{equation}
\Big(\nabla^2 -\frac{\partial^2}{\partial t^2}\Big) {\bf E_1} = -3 C  e^{-3i\omega t}  e^{-3 \rho^2 /w_0^2} \Big(e^{ikz}+
r e^{-ikz}\Big){\bf e_x}
\end{equation}
\begin{equation}
\Big(\nabla^2 -\frac{\partial^2}{\partial t^2}\Big){\bf B_1} = - C e^{-3i\omega t}  e^{-3 \rho^2 /w_0^2} \Big(e^{ikz}-
r e^{-ikz}\Big){\bf e_y},
\end{equation}
where $C=  16 r \omega^2 \mathcal{E}_0^3$.

It then follows that we can set
$$
{\bf E_1} = E_1 {\bf e_x} , \quad {\bf B_1} = B_1
{\bf e_y},
$$
and using the retarded Green function, the solution of the above equations for the electric field turns out to be
\begin{equation}
E_1=  \frac{3}{4\pi}C e^{-3i\omega t} \int d\rho' ~dz' ~\rho' ~d\phi'~e^{-3 (\rho' /w_0)^2} \Big(e^{ikz'}+ r e^{-ikz'}\Big)
\frac{e^{3i \omega |{\bf x}-{\bf x'}|}}{
{|{\bf x}-{\bf x'}|}}.
\end{equation}

Now, if the beam is highly concentrated around the axis $\rho'=0$, we can approximate
$$
|{\bf x}-{\bf x'}| \approx [\rho^2 + (z-z')^2]^{1/2},
$$
and the above integral reduces to
\begin{equation}
E_1=   \frac{C}{4} w_0^2 e^{-3i\omega t} (1-e^{-3(R/w_0)^2}) (I_+ + r I_- ) ,\label{E1}
\end{equation}
where
\begin{equation}
I_{\pm} (\rho , z)=
 \int_0^L dz'  e^{\pm ikz'}
\frac{e^{3i \omega [\rho^2 + (z-z')^2]^{1/2}}}{
{[\rho^2 + (z-z')^2]^{1/2}}},\label{II}
\end{equation}
$R$ is the radius of the interferometer and $L$ its length.

For the magnetic field $B_1$, just change $r \rightarrow -r$ in \eqref{E1} and divide by a factor 3.

As shown in the appendix below, the integrals \eqref{II} can be solved approximately. Thus, the corrections to the EM field due to non-linear effects can be approximated as
$$
\Delta E_x=  4\lambda \mathcal{E}_0^3  r \omega^2 w_0^2 \Big(1-e^{-3(R/w_0)^2}  \Big)      e^{-3i\omega t +i\sqrt{8} ~k\rho  +\pi i /4} \sqrt{\frac{\pi}{\sqrt{2}~k\rho}}~
$$
\begin{equation}
\times \Big[       e^{ikz }  ~ \Theta (z-\rho /\sqrt{8})
+r e^{ -ikz}  ~  \Theta (L-\rho /\sqrt{8}- z) \Big] \label{EEE},
\end{equation}
($\Theta$ is the step-function). The same formula for $B_y$ applies with  the only changes $r \rightarrow -r$ in the term between square brackets in Eq. \eqref{EEE} and an overall factor 1/3.

The above equation \eqref{EEE} can be written in general units as
\begin{equation}
\frac{|\Delta E_x|}{\mathcal{E}_0} = \frac{4\pi}{45 } \Big(\frac{r_0}{R}\Big)^2   \Big(\frac{w_0}{\lambda_L}\Big)\Big(1-e^{-3(R/w_0)^2}  \Big) \sqrt{\frac{\lambda_L}{2\sqrt{2}~\rho}}
 \Big( \frac{\mathcal{P}}{\mathcal{P}_e}\Big) ,\label{P}
\end{equation}
where $r_0=\alpha \hbar /m_ec \approx 2.8 \times 10^{-15}$ m is the classical electron radius,
$$\mathcal{P}_e \equiv  m_e^2 c^4/\hbar  \approx 6.7 \times 10^7 \textrm{W}
$$
 is the basic ``quantum power'' associated to the electron, $\mathcal{P}=\pi R^2 \mathcal{E}_0^2$ is the laser power and $\lambda_L$ is its wavelength.
The  formula \eqref{P}  is written as a dimensionless quantity in order to give an idea of the order of magnitude of the nonlinear effects.

\section{Radiation pressure}

With the above results, we can now calculate the contribution of the non-linear electrodynamic effects to the intensity as given by the Poynting vector
(which is entirely in the $z$ direction):
\begin{equation}
\mathcal{I}(\rho, z)= \frac{1}{2} | {\bf E} \times {\bf B^*} + {\bf E^*} \times {\bf B} |.
\end{equation}

It is enough to restrict the calculations to the end points of the interferometer and to consider $L>\rho/8$. To order $\lambda$, the Poynting vector oscillates with a period $2\omega$, and therefore its time average makes no contribution to the radiation pressure on the average. For the time averaged effect, we  must go to order $\lambda^2$.

Let us first consider the EM field at the end of the interferometer, $z=L$. There, only the right moving wave contributes to the QED correction. Accordingly, the contribution to the Poynting vector from the right-moving wave turns out to be:
\begin{equation}
\langle\Delta \mathcal{I}(\rho, L)\rangle= \frac{16 \sqrt{2}}{3 }\pi^2 \lambda^2 |\mathcal{E}_0|^6 |r|^2  ~ \omega^3 w_o^4 \Big(1-e^{-3(R/w_0)^2}  \Big)^2 R,\label{<>}
\end{equation}
after an integration over the cross section of the beam (i.e., $0<\rho < R$).

Likewise,
\begin{equation}
\langle\Delta \mathcal{I}(\rho, 0)\rangle=|r|^2
\langle\Delta \mathcal{I}(\rho, L)\rangle.
\end{equation}

Formula \eqref{<>} can be written in general units as
\begin{equation}
\langle\Delta \mathcal{I}(\rho, L)\rangle= \big(\frac{2 \mathcal{P}}{c}\Big) \Big[
\frac{64\sqrt{2}}{3(45)^2 } |r|^2 \Big(1-e^{-3(R/w_0)^2}  \Big)^2 \Big(\frac{r_0}{R}\Big)^4 ~ \frac{w_0^4}{ \lambda_L^3 R}~\Big(\frac{\mathcal{P}}{\mathcal{P}_e}\Big)^2 \Big].\label{hbar}
\end{equation}

The \emph{force} on the mirror at one end of the interferometer is precisely $\langle\Delta \mathcal{I}(\rho, L)\rangle$, to be compared with the force  $2 \mathcal{P}/c$ produced by the classical radiation pressure  (see e.g., \cite{magg}, Sect. 9.4.2). In Eq. \eqref{hbar}, the term in square brackets is precisely the dimensionless QED correction to the radiation pressure force.

\section{Conclusions}

The formulas obtained above describe the generation of waves of energy $3\omega$ with a propagation vector along a conical direction $\sqrt{8} ~{\bf e}_{\rho} \pm {\bf e}_z$,  with a opening semi-angle $ \alpha =\arccos (1/3) \approx 70^{\circ} $. One can interpret this result as describing the splitting along a conical direction of two beams of  photons colliding head-on. However, it must be noticed  that the electric and magnetic field vectors remain in the  $z$ constant plane and the Poynting vector is in the $z$ direction: this can be interpreted as a sideways propagation of the generated waves (as in anisotropic crystals).

In the particular case of an extremely powerful laser, of about 750 kW as reached  by LIGO \cite{ligo} (using a dual-recycled Fabry-P\'erot-Michelson interferometer \cite{ligo2012}), and a wave length $\lambda_L \approx 1000$ nm., with $R \approx w_0 \approx 10$ cm, the above correction factor, as given by \eqref{hbar}, turns out to be of the order of  $10^{-33}$. Clearly, it can be safely neglected in all measurements. There remains, nevertheless, the possibility that the effects of nonlinear electrodynamic could be detected in the future with extremely energetic light beams.

\section*{Appendix}

\renewcommand{\theequation}{\textrm{A}.\arabic{equation}}

Consider  $I_+(\rho , z)$ first. Notice that the integrand oscillates very rapidly, except around the minimum of the function
$$
w(z') \equiv  z' + 3 [\rho^2 + ({z-z'}^2)]^{1/2}
$$
appearing  in the exponent. This minimum is located at $z'_0 = z- \rho/\sqrt{8}$ which falls inside the integration range if $0 < z'_0  < L$ and outside it otherwise. Accordingly, the main contribution to $I_+(\rho,z)$ comes from the point $z_0'$ in the former case, while  $I_+(\rho,z)$ is negligible in the latter case. Expanding $w(z')$ around its minimum,
$$
w(z')= z + \sqrt{8} \rho  +\frac{8\sqrt{2}}{9\rho}~(z'-z+\rho/ \sqrt{8})^2 +...
$$
and extending, without much error, the limits of integration of the above integral to infinity, one finds as a rough approximation:
\begin{equation}
 I_+ (\rho , z) \approx \sqrt{\frac{\pi}{\sqrt{2} ~k \rho}}~e^{ikz +i \sqrt{8} k \rho + \pi i /4} ~\Theta(z-\rho/ \sqrt{8} ),
\end{equation}
where $\Theta$ is the step-function ($\Theta (x) =1$ if $x>0$ and $0$ is $x<0$).

An entirely similar argument leads to
\begin{equation}
 I_- (\rho , z) \approx \sqrt{\frac{\pi}{\sqrt{2} ~k \rho}}~e^{-ikz +i \sqrt{8} k \rho + \pi i /4} ~\Theta(L -\rho/ \sqrt{8} -z).
\end{equation}

\end{document}